\documentclass[prb,showpacs,preprintnumbers,amsmath,amssymb,twocolumn]{revtex4-1}
\usepackage{graphicx}
\usepackage{dcolumn}
\usepackage{bm}
\usepackage{xcolor}

\usepackage[mathletters]{ucs}
\usepackage[utf8x]{inputenc}

\usepackage{latexsym}
\usepackage{graphicx}

\usepackage{subfigure}    
\usepackage{epsfig}

\begin{document}

\title{Uncovering the behavior of Hf$_2$Te$_2$P and the candidate Dirac metal Zr$_2$Te$_2$P}

\author{K. -W. Chen,$^{1,2}$ S. Das,$^{1,2}$ D. Rhodes,$^{1,2}$ S. Memaran,$^{1,2}$ T. Besara,$^1$ T. Siegrist,$^{1,3}$ E. Manousakis,$^{1,2}$ L. Balicas,$^1$ and R. E. Baumbach$^1$}
\affiliation{$^1$National High Magnetic Field Laboratory, Florida State University}
\affiliation{$^2$Department of Physics, Florida State University}
\affiliation{$^3$Department of Chemical and Biomedical Engineering, Florida State University}
\date{\today}

\begin{abstract}
Results are reported for single crystal specimens of Hf$_2$Te$_2$P and compared to its structural analogue Zr$_2$Te$_2$P, which was recently proposed to be a potential reservoir for Dirac physics.[\onlinecite{Huiwen2015}] Both materials are produced using the iodine vapor phase transport method and the resulting crystals are exfoliable. The bulk electrical transport and thermodynamic properties indicate Fermi liquid behavior at low temperature for both compounds. Quantum oscillations are observed in magnetization measurements for fields applied parallel but not perpendicular to the $c$-axis, suggesting that the Fermi surfaces are quasi-two dimensional. Frequencies are determined from quantum oscillations for several parts of the Fermi surfaces. Lifshitz-Kosevich fits to the temperature dependent amplitudes of the oscillations reveal small effective masses, with a particularly small value $m^*$ $=$ 0.046$m_0$ for the $\alpha$ branch of Zr$_2$Te$_2$P. Electronic structure calculations are in good agreement with quantum oscillation results and illustrate the effect of a stronger spin-orbit interaction going from Zr to Hf. These results suggest that by using appropriate tuning parameters this class of materials may deepen the pool of novel Dirac phenomena.
\end{abstract}

\maketitle
\section{Introduction}
Efforts in condensed matter physics during the past century have focused on states of matter that are characterized by their symmetry. This encompasses phenomena such as magnetism, superconductivity, charge density wave order, nematicity, etc. and was foundational for development of society transforming modern technologies: e.g., those based on rare earth magnets and superconductors. Another type of ordering has recently gained prominence, where the electronic state is characterized by its topology instead of a broken symmetry. The first examples of this were the quantum Hall effect and fractional quantum Hall effect (e.g., in GaAs/AlGaAs)~\cite{Klitzing1986, Stormer1999} and during the past decade this field has rapidly expanded to include new families of topological materials, particularly those with three dimensional crystalline structures. These three dimensional Dirac materials host topologically protected linearly dispersing (“photon-like”) electronic bands both in the bulk and as surface states.~\cite{wehling2014} There are various origins for this behavior including strong spin-orbit coupling (e.g., topological insulators Bi$_2$Te$_3$ \cite{Zhang2009, Chen178, Hsieh2009} and Bi$_2$Se$_3$ \cite{Zhang2009, Xia2009, Hsieh2009}), crystalline symmetries (e.g., Dirac semimetals Cd$_3$As$_2$ \cite{Liu2014, Wang2013} and Na$_3$Bi \cite{Wang2012, Liu2014a}), combinations of these effects (Weyl semimetals (Nb,Ta)(P,As) \cite{Weng2015, Lv2015, Xu2015, Shekhar2015}), and strong electronic correlations (Kondo insulators SmB$_6$ \cite{Dzero2010, Xu2013, Jiang2013}). 

Given their novel physics and potential practical applications,\cite{wehling2014,Liang2015,Shekhar2015} there is great interest in diversifying the pool of 3D Dirac materials. Taking inspiration from earlier successes in condensed matter physics, this will be accomplished largely through tuning of known systems (and new materials as they are uncovered) using control parameters such as magnetic fields, applied pressure, chemical substitution, nanostructuring, and device fabrication. This route not only gives insight into parent compounds, but also has a high likelihood of uncovering transformative phenomena. This approach has already proven successful for some topological systems: in Cd$_3$As$_2$ both applied pressure \cite{He2015} and point contact experiments \cite{Aggarwal2016} stabilize superconductivity and may provide a route towards studies of Majorana fermions. 

Zr$_2$Te$_2$P was recently introduced as a possible strongly topological metal with multiple Dirac cones: i.e., a Dirac metal when exfoliated into a few atomic layers.\cite{Huiwen2015} Of particular interest is that this compound crystallizes in the well-known tetradamite structure and is easily exfoliated, opening the way for device development. In order to explore the effect of a strengthened spin-orbit interaction in this system, we undertook to synthesize related compounds with larger Z elements. Here we report the synthesis and bulk characterization of the isoelectronic analogue Hf$_2$Te$_2$P, which forms in the same structure, is easily exfoliated, and supports similar bulk electrical transport and thermodynamic behavior. Quantum oscillations are observed in both compounds starting near magnetic fields of $H$ = 3 T, opening the possibility of experimentally comparing their Fermi surfaces and characterizing the Dirac behavior through the Berry’s phase. We additionally present electronic structure calculations which reveal the effect of a strengthening spin-orbit interaction going from Zr to Hf.

\section{Methods}
Single crystals of Zr$_2$Te$_2$P and Hf$_2$Te$_2$P were grown from elements as previously described for the Zr version.\cite{Tschulik2009} Stoichiometric polycrystalline precursor material was first prepared by reacting the raw elements at 1000 $^o$C for 24 hours. The resulting powders (1.5 g and 1.3 g for Zr$_2$Te$_2$P and Hf$_2$Te$_2$P, respectively) were subsequently sealed under vacuum with 46 mg of iodine in quartz tubes with 18 mm diameter, 2 mm thickness, and 10 cm length. The tubes were placed in a resistive tube furnace with a temperature gradient with hot and cold zone temperatures of 900 and 800 $^o$C, respectively. The ampoules were held under these conditions for three weeks and then quenched to room temperature. Large hexagonal single crystals formed near the cold zone. For Zr$_2$Te$_2$P the linear dimensiona are as large as 10 $\times$ 10 $\times$ 0.4 mm. Hf$_2$Te$_2$P crystals were slightly smaller, with dimensions 3 $\times$ 3 $\times$ 1 mm (Fig.~\ref{fig:XRD}). 

X-ray diffraction measurements were collected using a Scintag PAD-V $\Theta$:2$\Theta$ diffractometer using Cu K-alpha radiation. The sample consisted of crushed crystals spread randomly on a glass slide. Due to the tendency of the crystals to separate as plates in the $ab$-plane, completely random orientation was not achieved, and the XRD pattern shows signs (001) texture. The pattern was fit with the WINPREP software~\cite{stahl} using the known parameters of the isostructural compound Zr$_2$Te$_2$P as starting parameters.

Flakes of Zr$_2$Te$_2$P and Hf$_2$Te$_2$P were exfoliated from bulk crystals using the micromechanical cleavage technique with adhesive scotch-tape and transferred onto a $p$-doped silicon wafer substrate covered with a 270 nm layer of SiO$_2$. Atomically-thin crystals were identified under an optical microscope using optical contrast. Atomic Force Microscopy (AFM) imaging was performed using an Asylum Research MFP-3D AFM system. Images were taken about 2 hours after exfoliation, during which the samples were exposed to air.

Zero magnetic field electrical resistivity $\rho$ was measured using the $^3$He option in Quantum Design Physical Properties Measurement System for temperatures 400 mK $<$ $T$ $<$ 300 K. Several individual crystals were measured for each concentration, which revealed a high degree of batch uniformity. Heat capacity $C$ measurements were performed using the same apparatus for temperatures 400 mK $<$ $T$ $<$ 20 K. Magnetization $M(T,H)$ measurements were undertaken for single crystals for temperatures $T$ $=$ 1.8 - 300 K under an applied magnetic field of $H$ $=$ 10 kOe applied parallel $\parallel$ and perpendicular $\perp$ to the $c$-axis using a Quantum Design Magnetic Property Measurement System. Magnetic susceptibility $\chi$ is defined as the ratio $M/H$. 

Electronic structure calculations were performed using the Vienna \emph{ab-initio} simulation package~\cite{Shiskin_PRL_2007, Fuchs_PRB_2007, Shiskin_PRB_2007, Shiskin_PRB_2006} (VASP) within the generalized gradient approximation (GGA). The contribution from the spin-orbit coupling is included in the calculations. The Perdew-Burke-Ernzerhof (PBE) exchange correlation functional ~\cite{Perdew_PRL_1996} and the projected augmented wave (PAW) methodology~\cite{Blochl_PRB_1994} were used to describe the core electrons. The 4\emph{s}, 4\emph{p}, 5\emph{s}, and 4\emph{d} electrons 
for Zr, the 
5\emph{s}, 5\emph{p},  6\emph{s}, and 5\emph{d}  electrons for Hf,  the 4\emph{d}, 5\emph{s} and 5\emph{p} electrons for Te, and the 3\emph{s} and 3\emph{p} electrons 
for P were treated as valence electrons 
in all calculations. The energy cut off for the plane-wave basis was chosen to be 400 meV. A total of 240 bands and a $k-$point mesh of $20\times20\times4$ were used for the self-consistent ground state calculations. A total of 100 $k-$points were chosen between each pair of special $k-$points in the Brillouin-zone for the band-structure calculations. The Fermi surfaces were generated with a $k-$point mesh of $20\times20\times4$, using the eigenvalues obtained from VASP and were visualized using the XCrysden software.~\cite{Kokalj_CMS_2003}

\section{Results and Discussion}

\begin{figure*}[!tht]
    \begin{center}
        \includegraphics[width=5in]{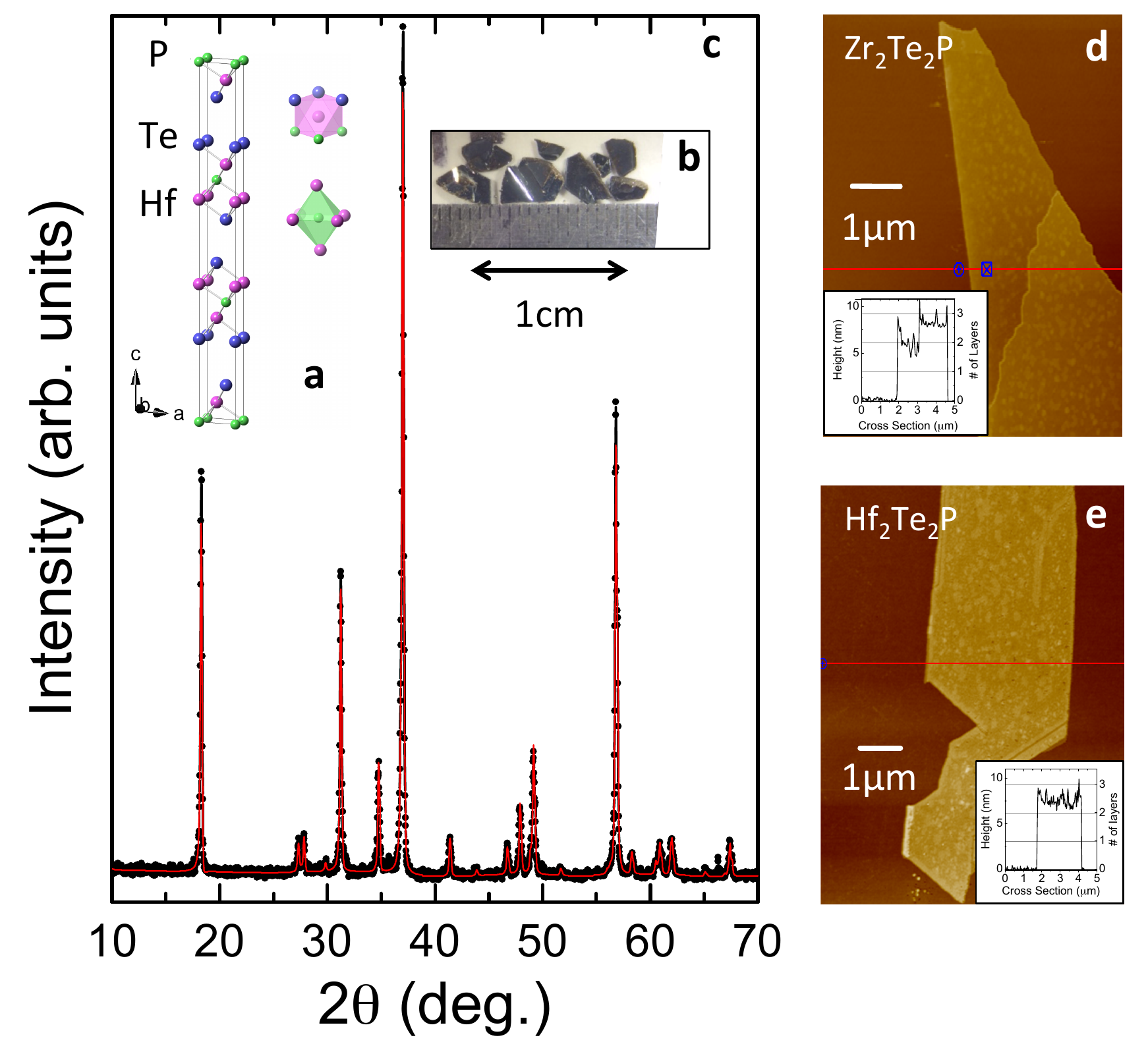}
        \caption{(a) Unit cell of the $M$$_2$Te$_2$P ($M$ $=$ Zr and Hf) compounds. (b) Single crystal specimens of Hf$_2$Te$_2$P. (c) Powder x-ray diffraction pattern for Hf$_2$Te$_2$P. The solid red line is a patternf fit to the data to determine lattice parameters. (d) Atomic Force Microscopy (AFM) image collected from an atomically thin flake of Zr$_2$Te$_2$P on a SiO$_2$ substrate. Inset: Height Profile (along the red line shown in the image) indicating a thickness of approximately 6nm equivalent to two unit cells. (e) AFM image for Hf$_2$Te$_2$P. Inset: Height profile for two-unit cell flake.}
        \label{fig:XRD}
    \end{center}
\end{figure*}

Results from powder X-ray diffraction measurement of crushed Hf$_2$Te$_2$P crystals are shown in Fig.~\ref{fig:XRD}, along with a pattern fit. Parameters of the isostructural compound Zr$_2$Te$_2$P, which crystallizes in the space group R$\bar{3}$m (\#166), were used as starting parameters. The fit yielded lattice parameters $a$ $=$ 3.7946(2) \AA~and $c$ $=$ 29.140(1) \AA, and displayed no traces of impurity phases. The lattice parameters are smaller than the Zr$_2$Te$_2$P parameters ($a$ $=$ 3.8119(3) \AA~ and $c$ $=$ 29.189(3) \AA),\cite{Tschulik2009} indicative of the smaller ionic radius of Hf compared to Zr.~\cite{teatum}

Atomic force microscopy (AFM) results for exfoliated flakes of Zr$_2$Te$_2$P and Hf$_2$Te$_2$P are shown in Figs.~\ref{fig:XRD}d,e. Edge step profiles reveal that the crystals are easily reduced to single- or several-unit cell flakes, where the edge step can be the same size as a single $c$-axis unit cell. Samples that are left exposed to air are seen to change over time, revealing that these specimens are air sensitive. 

\begin{figure*}[!tht]
    \begin{center}
        \includegraphics[width=5in]{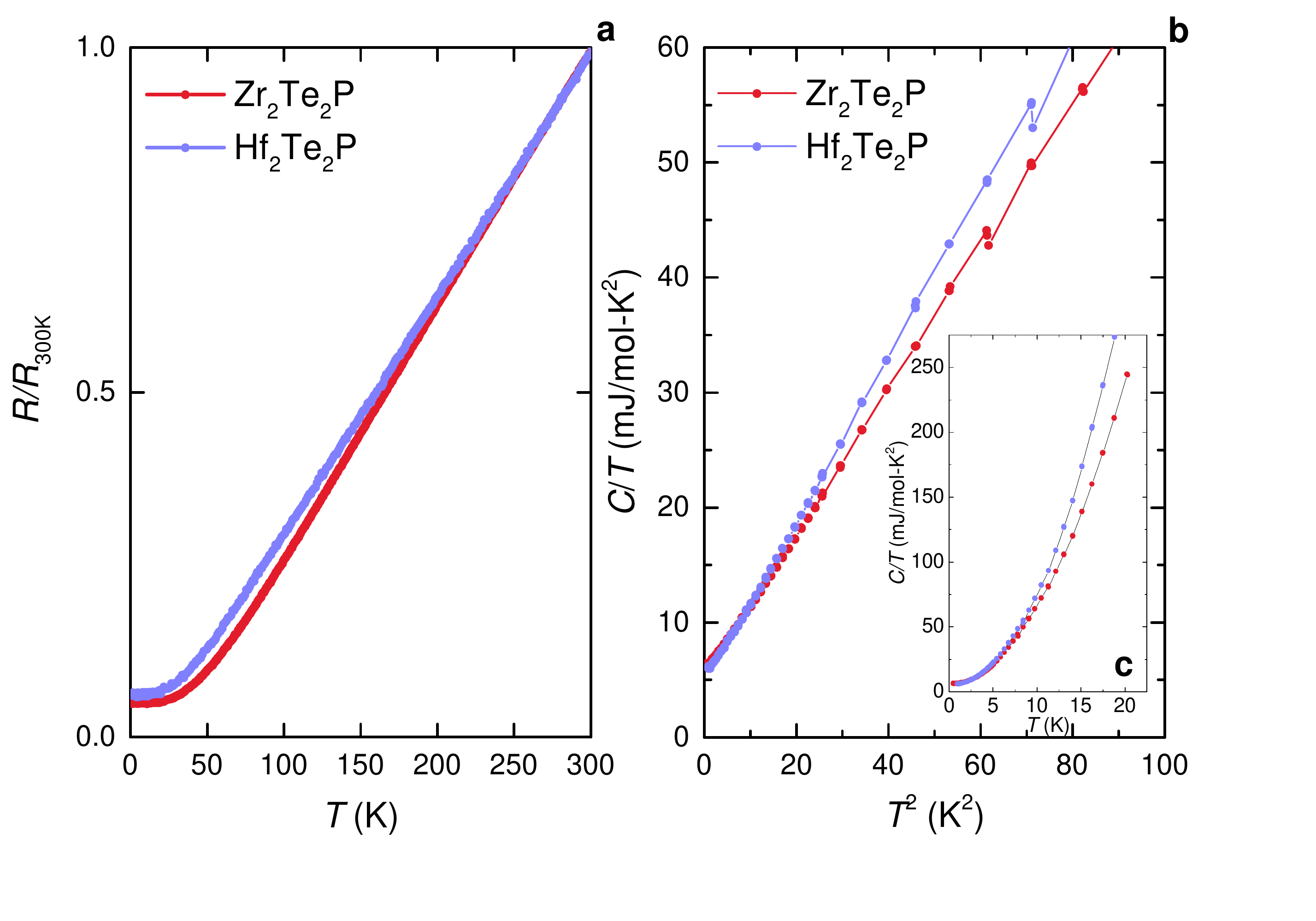}
        \caption{(a)Room temperature normalized electrical resistance $R/R_{\rm{300K}}$ for $M$$_2$Te$_2$P ($M$ $=$ Zr and Hf). (b) Heat capacity $C$ divided by temperature $T$ vs. $T^2$ for $M$$_2$Te$_2$P ($M$ $=$ Zr and Hf) showing linear behavior as expected for a Fermi liquid. (c) $C/T$ vs $T$ for $M_2$Te$_2$P ($M$ = Zr and Hf.)}
        \label{fig:rhoHC}
    \end{center}
\end{figure*}

Electrical resistance normalized to its room temperature value $R/R_{\rm{300K}}$ as a function of temperature $T$ data are shown in Fig.~\ref{fig:rhoHC}a for Zr$_2$Te$_2$P and Hf$_2$Te$_2$P, where the electrical current was applied in the basal $ab$-plane. Metallic behavior is observed for both materials, with room temperature resistivities $\rho$$_{\rm{300K}}$ $\approx$ 60 $-$ 80 $\mu$$\Omega$cm, depending on uncertainty of the geometric factor. $R/R_{\rm{300K}}$ decreases linearly with decreasing temperature down to $T$ $\approx$ 50 K, where the resistance begins to saturate in a Fermi-liquid like manner. For both compounds, the resistivity saturates near 3 $-$ 5 $\mu$$\Omega$cm, giving a residual resistivity ratio $RRR$ = $R_{\rm{300K}}$/$R_{\rm{0}}$ $\approx$ 20. This value for $RRR$ is not particularly large, and might be taken to indicate appreciable disorder scattering. Alternatively, it is possible that as-cast crystals have large $RRR$ but rapidly begin to degrade after being exposed to atmosphere. As shown below, quantum oscillations are easily observed in magnetization measurements, which is unexpected if there is large disorder scattering. 

The temperature dependences of the heat capacity $C$ divided by $T$ are shown in Figs.~\ref{fig:rhoHC}b,c for Zr$_2$Te$_2$P and Hf$_2$Te$_2$P. Fermi liquid behavior is seen for both compounds, where the expression $C/T$ $=$ $\gamma$ $+$ $\beta$$T^2$ describes the data for 500 mK $<$ $T$ $<$ 10 K. Fits to the data give $\gamma$ $=$ 5.4 mJ/mol-K$^2$ and 4.7 mJ/mol-K$^2$ and $\beta$ $=$ 0.62 and 0.71 mJ/mol-K$^4$ for Zr$_2$Te$_2$P and Hf$_2$Te$_2$P, respectively. That the electronic coefficients of the heat capacity $\gamma$ are similar for both compounds is consistent with electronic structure calculations presented below, where the total density of states at the Fermi energy remains roughly constant upon going from Zr to Hf. From the expression $\beta$ $=$ $r$1944($T/\theta_{\rm{D}}$)J/mol-K, where $r$ is the number of atoms per formula unit, the Debye temperatures $\theta_{\rm{D}}$ are calculated to be 250 K and 239 K for the Zr and Hf compounds, respectively. This result is consistent with the increased atomic mass of Hf in comparison to Zr.

\begin{figure*}[!tht]
    \begin{center}
        \includegraphics[width=3.5in]{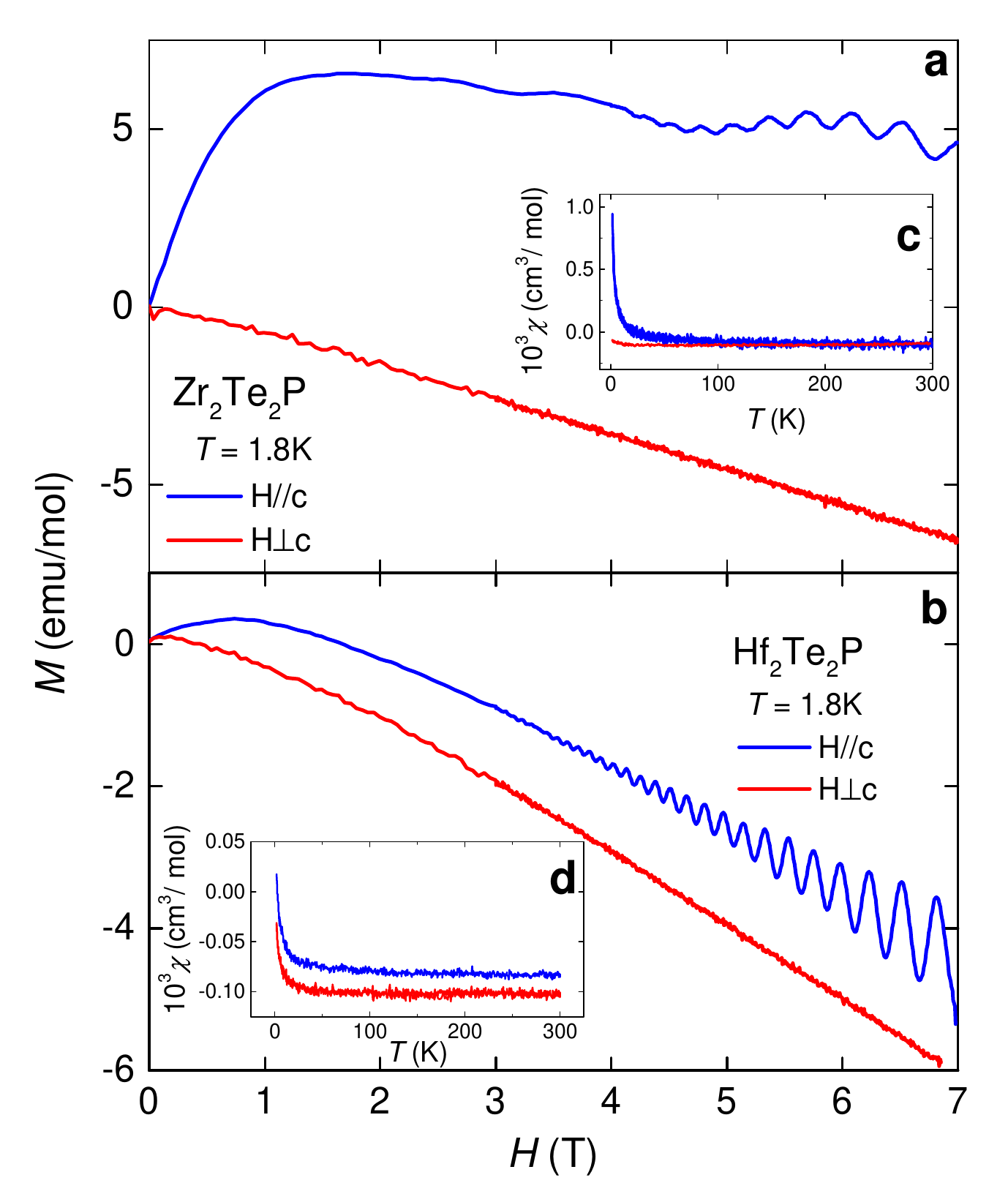}
        \caption{(a) Magnetization $M$ vs. magnetic field $H$ for temperature $T$ = 1.8 K for magnetic field $H$ applied parallel $\parallel$ and perpendicular $\perp$ to the $c$-axis for Zr$_2$Te$_2$P. (b) Magnetic susceptibility $\chi$ $=$ $M/H$ vs. $T$ for $H$ = 10 kOe applied $\parallel$ and $\perp$ to the $c$-axis for Zr$_2$Te$_2$P. (c) $M$ vs. $H$ for 1.8 K for $H$  $\parallel$ and $\perp$ to the $c$-axis for Hf$_2$Te$_2$P (d) $\chi$ vs. $T$ for $H$ = 10 kOe applied $\parallel$ and $\perp$ to the $c$-axis for Hf$_2$Te$_2$P.}
        \label{fig:M}
    \end{center}
\end{figure*}

Fig.~\ref{fig:M} displays magnetization $M$ as a function of the field $H$ applied either parallel $\parallel$ or perpendicular $\perp$ to the $c$-axis and for both compounds. $M(H)$ is diamagnetic for $H$ $\perp$ $c$, whereas curves for $H$ $\parallel$ $c$ initially are paramagnetic with increasing $H$, but eventually take on negative slopes, and even become negative for Hf$_2$Te$_2$P. The initial positive slopes can be attributed to a small fraction of paramagnetic impurities. In the insets, the magnetic susceptibilities $\chi$ $=$ $M/H$ for Zr$_2$Te$_2$P and Hf$_2$Te$_2$P are shown, where diamagnetic behavior with little anisotropy is observed at large temperatures. Weak upturns are seen at low $T$, which may be due to Curie tails from paramagnetic impurities. Quantum oscillations are observed starting from low fields for both compounds when $H$ $\parallel$ $c$. The very low field for onset of oscillations indicates the high sample quality. QOs are not seen when $H$ $\perp$ $c$, giving strong evidence that the Fermi surface is two dimensional. While this is expected from the band structure calculations presented below, further angle resolved measurements will be useful to illustrate the degree of two dimensionality. 

\begin{figure*}[!tht]
    \begin{center}
        \includegraphics[width=5in]{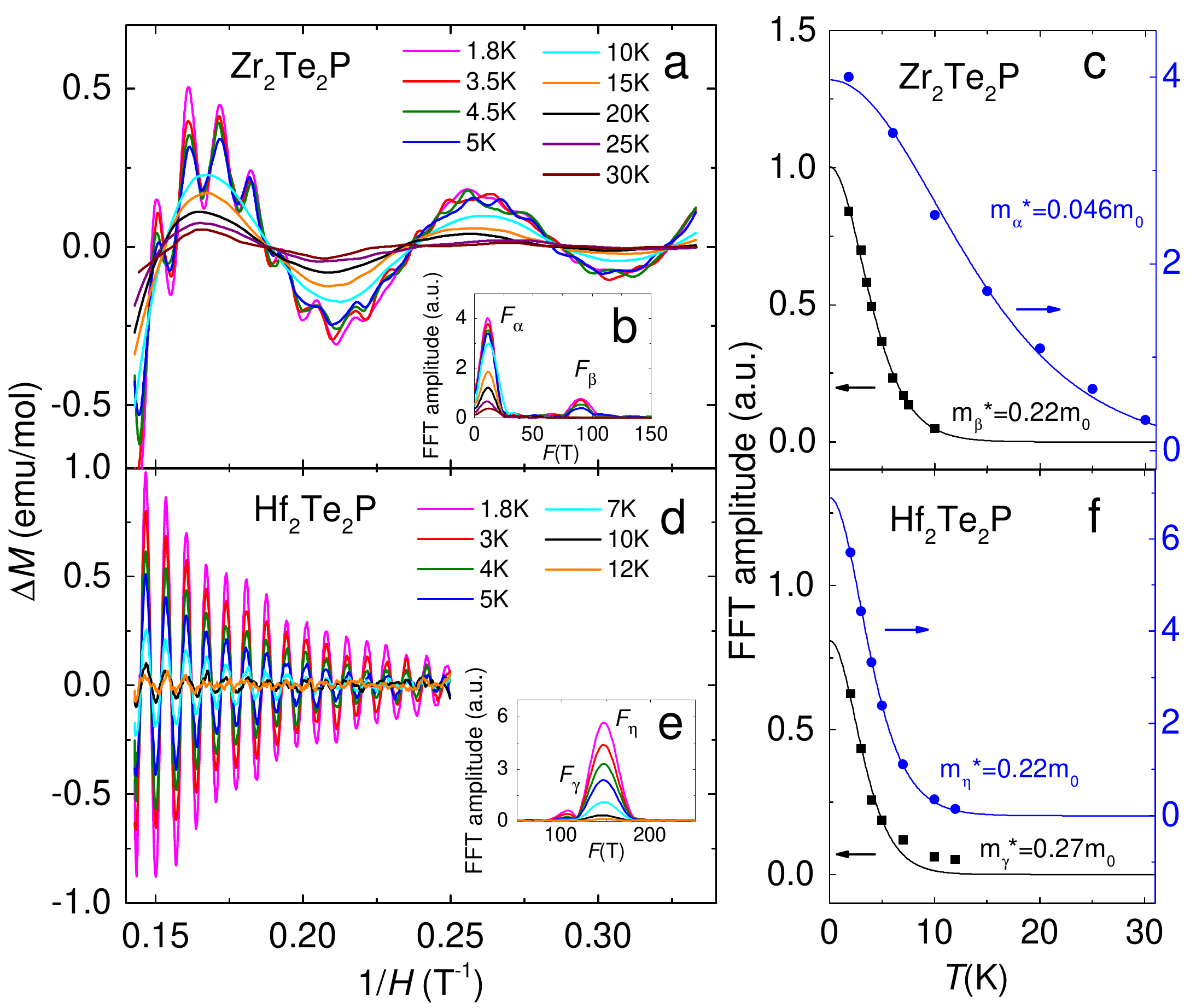}
        \caption{(a) Background subtracted magnetization $\Delta$$M$ for Zr$_2$Te$_2$P showing de Hass-van Alphen (dHvA) oscillations for different temperatures $T$ between 1.8 K and 30 K in a field range $H$ $=$ 3 T $-$ 7 T.  $\Delta$$M$ is obtained by subtracting a 3rd order polynomial background from the raw data. (b) Fast Fourier transforms (FFT ) for $\Delta$$M$ at various $T$s for Zr$_2$Te$_2$P. (c) Amplitude of the peaks observed in FFT spectra as a function of the temperature for Zr$_2$Te$_2$P. The solid lines are fits to the data using the Lifshitz-Kosevich formula. (d) $\Delta$$M$ for Hf$_2$Te$_2$P. (e) FFTs of $\Delta$$M$ at various $T$s (using a Hanning window) for Hf$_2$Te$_2$P. (f)  $T$ dependence of the FFT amplitude for $\Delta$$M$ for Hf$_2$Te$_2$P.}
        \label{fig:QO}
    \end{center}
\end{figure*}

The quantum oscillations in the magnetization $\Delta$$M$ are isolated by subtracting a 3rd order polynomial background as shown in Fig.~\ref{fig:QO} for different temperatures $T$ between 1.8 K and 30 K over the field range $H$ $=$ 3 T $-$ 7 T.  Fast Fourier transforms (FFT) for $\Delta$$M$ at various temperatures were performed using a Hanning window to extract the oscillation frequencies (Figs.~\ref{fig:QO}b,e). The FFTs yield frequencies $F_{\alpha}$ $=$ 11.4 T  and $F_{\beta}$ $=$ 89.9 T for Zr$_2$Te$_2$P and $F_{\gamma}$ $=$ 106.8 T  and $F_{\eta}$ $=$ 146.6 T for Hf$_2$Te$_2$P, as summarized in Table~\ref{tab:table1}. The temperature dependences of the FFT amplitude are shown in Figs.~\ref{fig:QO}c,f. The solid lines are fits to the data according to the thermal damping factor $R_{\rm{T}}$ of the Lifshitz-Kosevich formula using the expression,

\begin{equation}
R_{\rm{T}} = \frac{2\pi^2k_{\rm{B}}Tm^*/\hbar eH}{sinh(2\pi^2 k_{\rm{B}}Tm^*/\hbar eH)}
\label{eqn:1}
\end{equation}

where $m^*$ is the effective cyclotron mass. The fits yield small effective masses as collected in Table~\ref{tab:table1}. Particularly noteworthy is the very small $m^*$ $=$ 0.046$m_0$ for the $\alpha$ branch of Zr$_2$Te$_2$P at $F_{\rm{\alpha}}$ $=$ 11.4 T, where $m_0$ is the free electron mass.

\begin{table*}[!tht]
  \centering
  \caption{Summary of results from fast Fourier transform (FFT) of quantum oscillations for Zr$_2$Te$_2$P and Hf$_2$Te$_2$P. Frequencies $F$(T) and effective masses $m^*$($m_0$) are listed.}
  \label{tab:table1}
  \begin{tabular}{ |c|c|c|c|c|} 
 \hline
			   & $F_{\rm{\alpha}}$(T), $m_{\alpha}^*$($m_0$) & $F_{\rm{\beta}}$, $m_{\beta}^*$ & $F_{\rm{\gamma}}$, $m_{\gamma}^*$ & $F_{\rm{\eta}}$, $m_{\eta}^*$ \\ 
 Zr & 11.4, 0.046                               & 89.9, 0.22                            & ---                              & ---                           \\ 
 Hf & ---                               & ---                             & 106.8, 0.27                               & 146.6, 0.22                           \\ 
 \hline
\end{tabular}
\end{table*}

\begin{figure*}[htb]
    \begin{center}
        \subfigure[]{
            \includegraphics[width=0.45\textwidth]{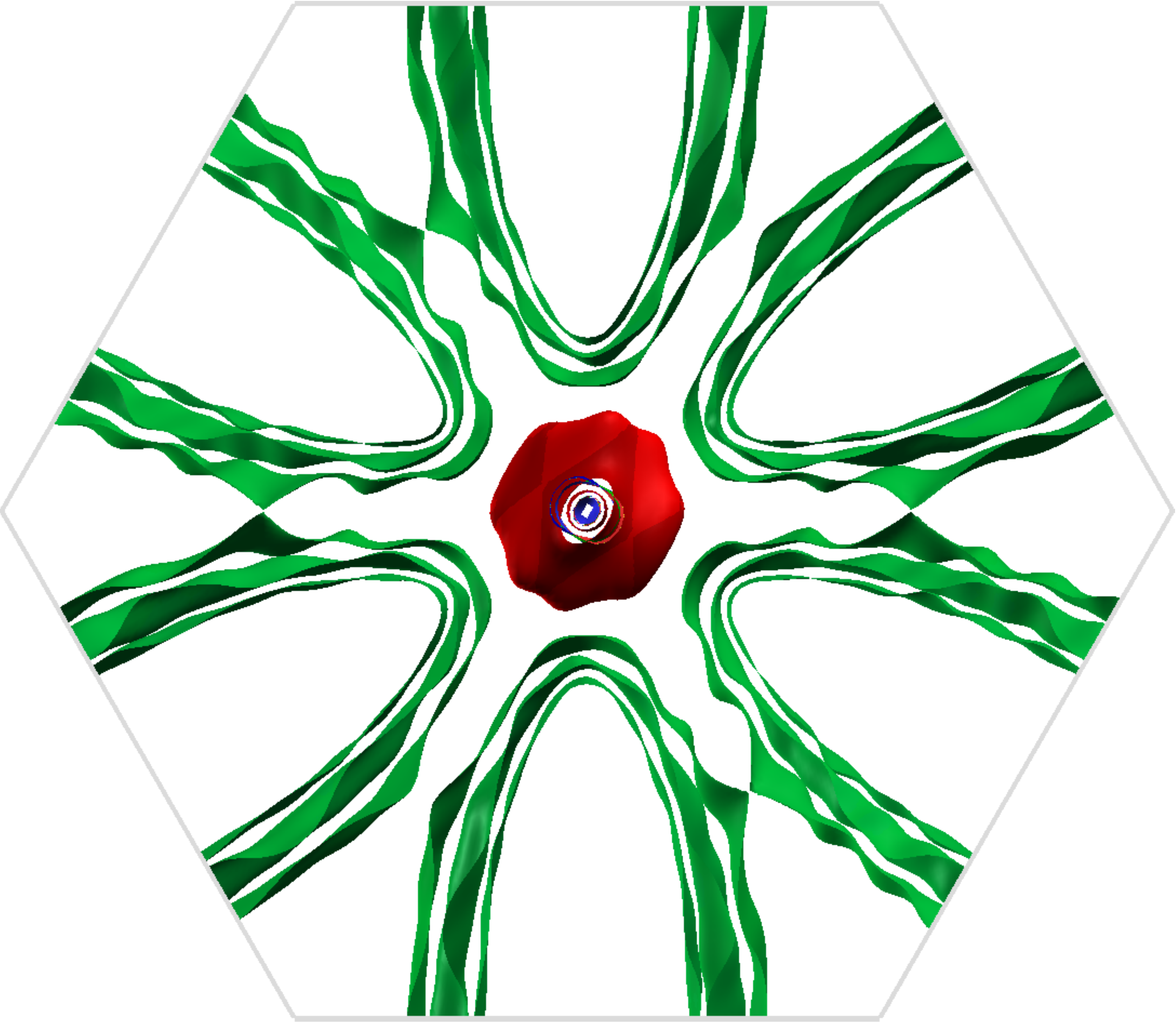}
            \label{fig:Zr_FS}
        } \hskip 0.3 in
        \subfigure[]{
            \includegraphics[width=0.45\textwidth]{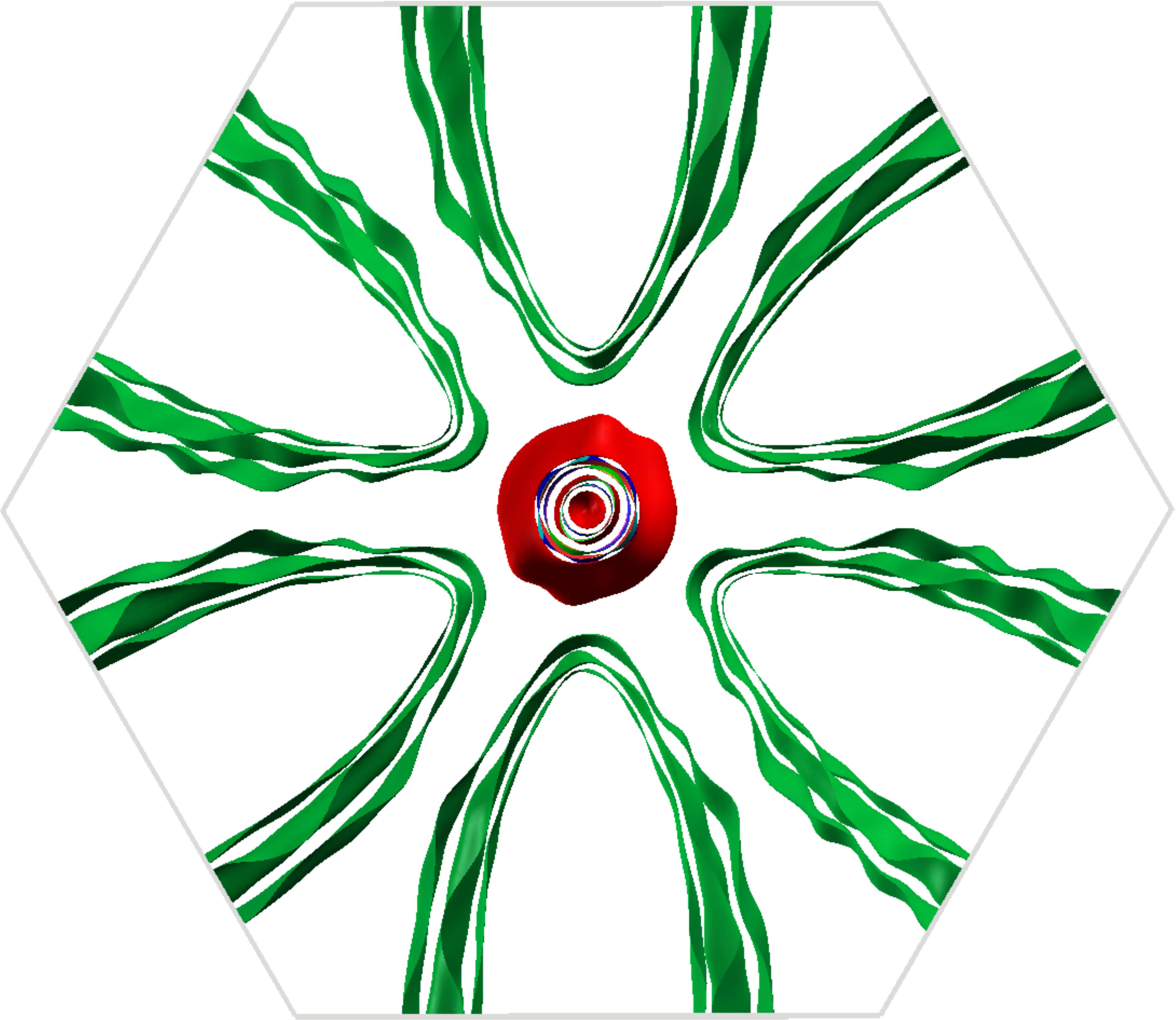}
            \label{fig:Hf_FS}
        }
    \end{center}
        \caption{(color-online) Comparison between the Fermi surface of the 
Zr (a) Hf (b) based 221 compounds. }
        \label{fig:FS}
\end{figure*}

\begin{figure*}[htb]
    \begin{center}
        \subfigure[]{
            \includegraphics[width=0.25\textwidth]{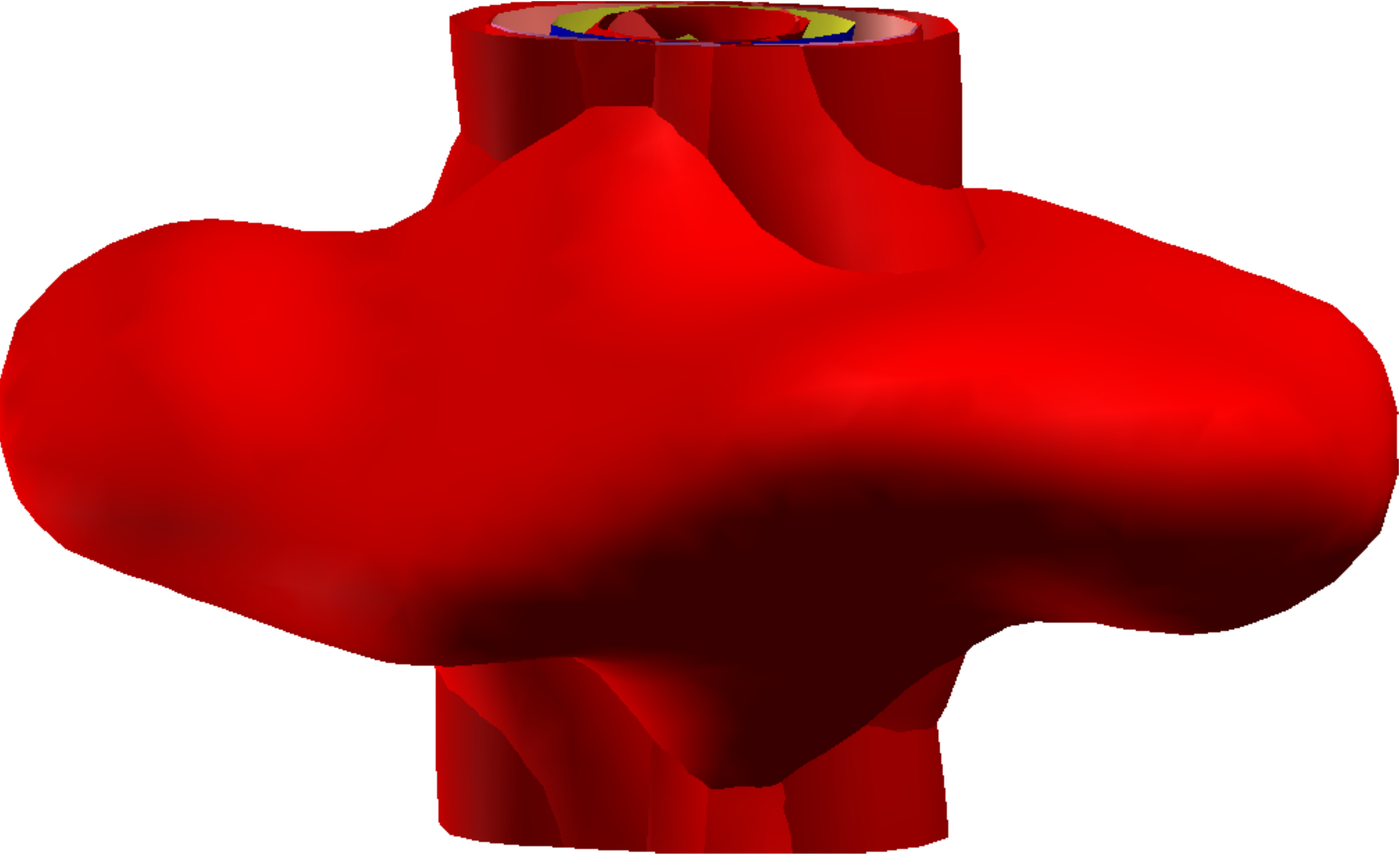}
            \label{fig:Zr_spin}
        } \hskip 0.6 in
        \subfigure[]{
            \includegraphics[width=0.25\textwidth]{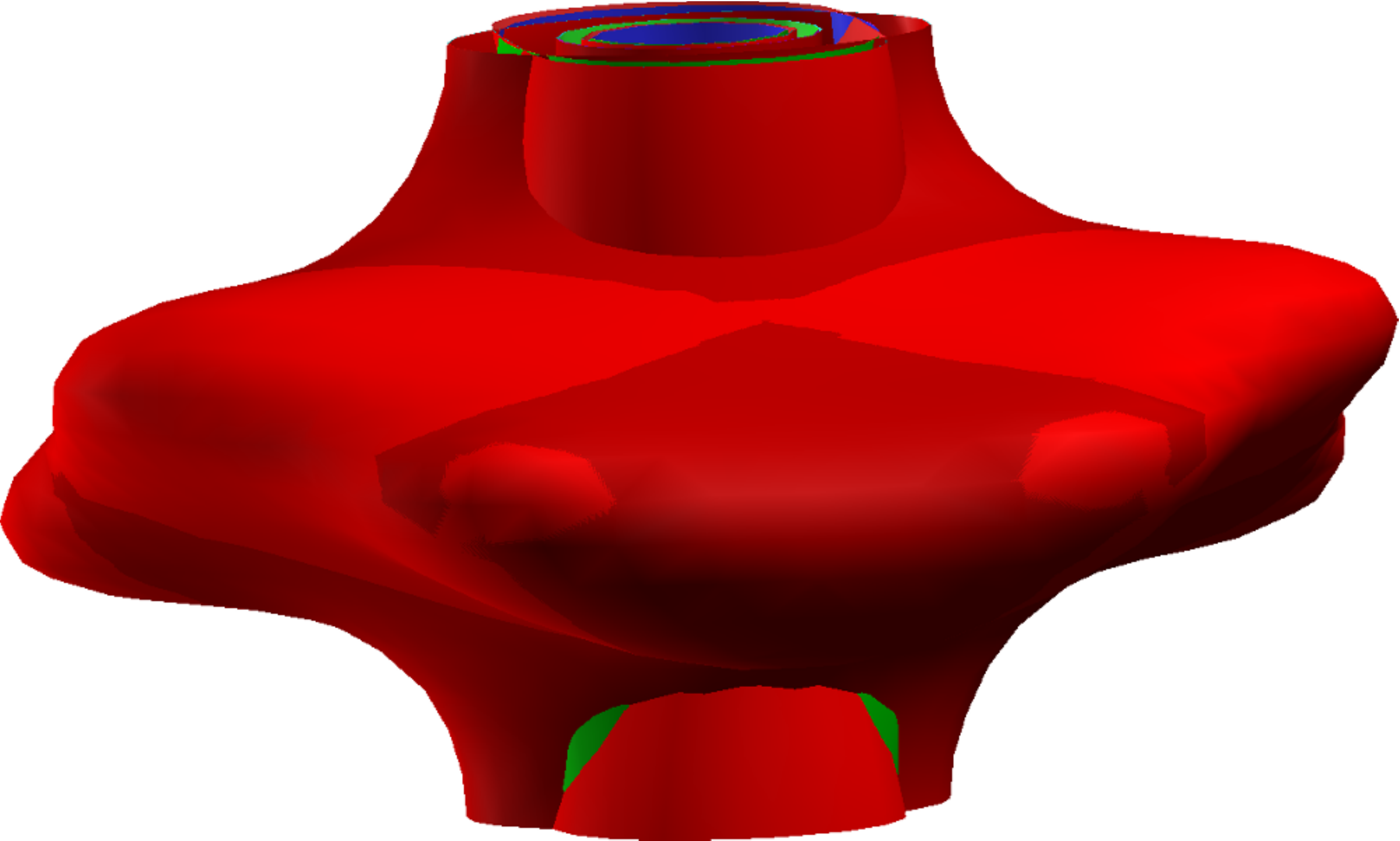}
            \label{fig:Zr_ls}
        }
        \\
        \subfigure[]{
            \includegraphics[width=0.25\textwidth]{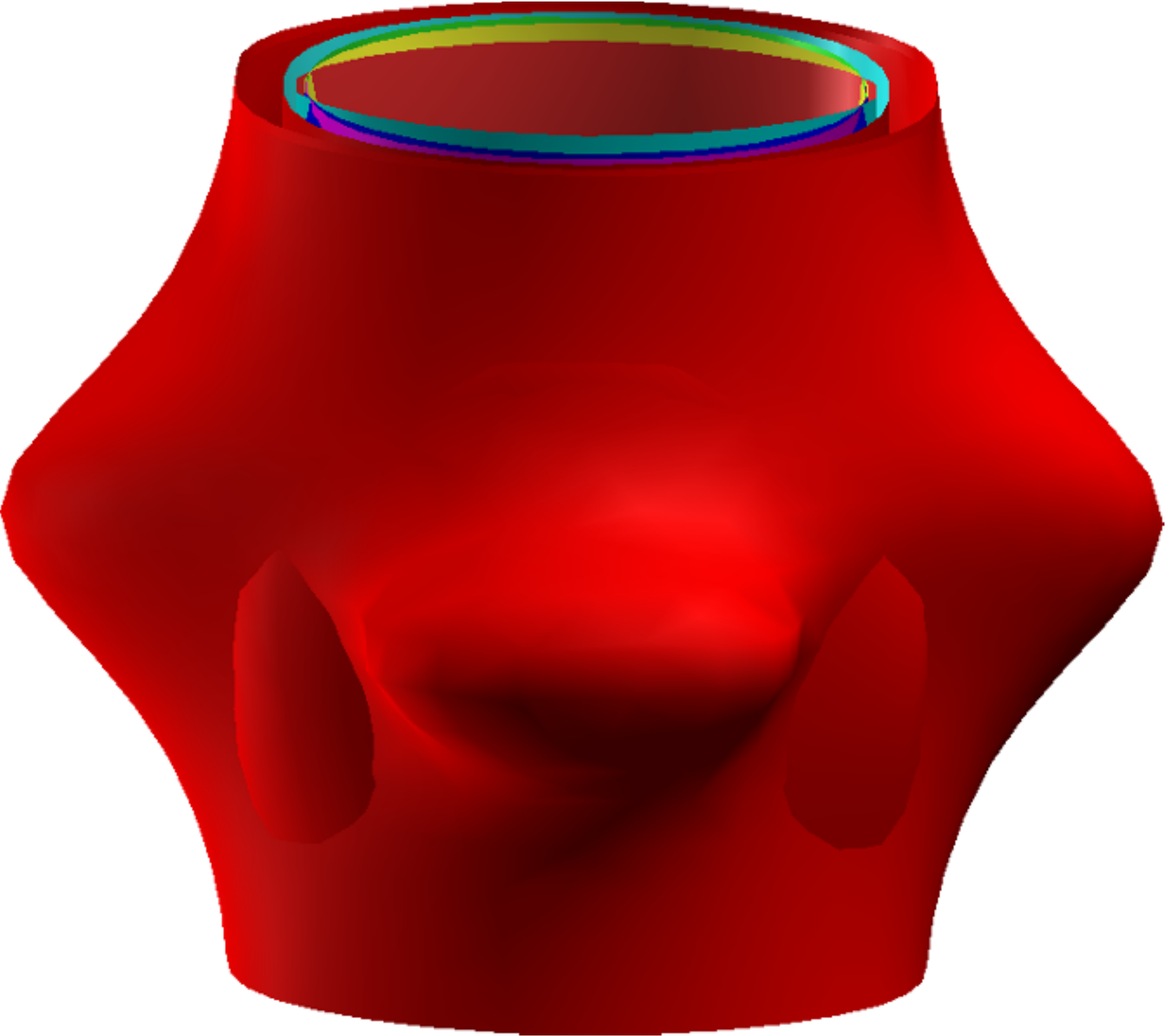}
            \label{fig:Hf_spin}
        } \hskip 0.6 in
        \subfigure[]{
            \includegraphics[width=0.25\textwidth]{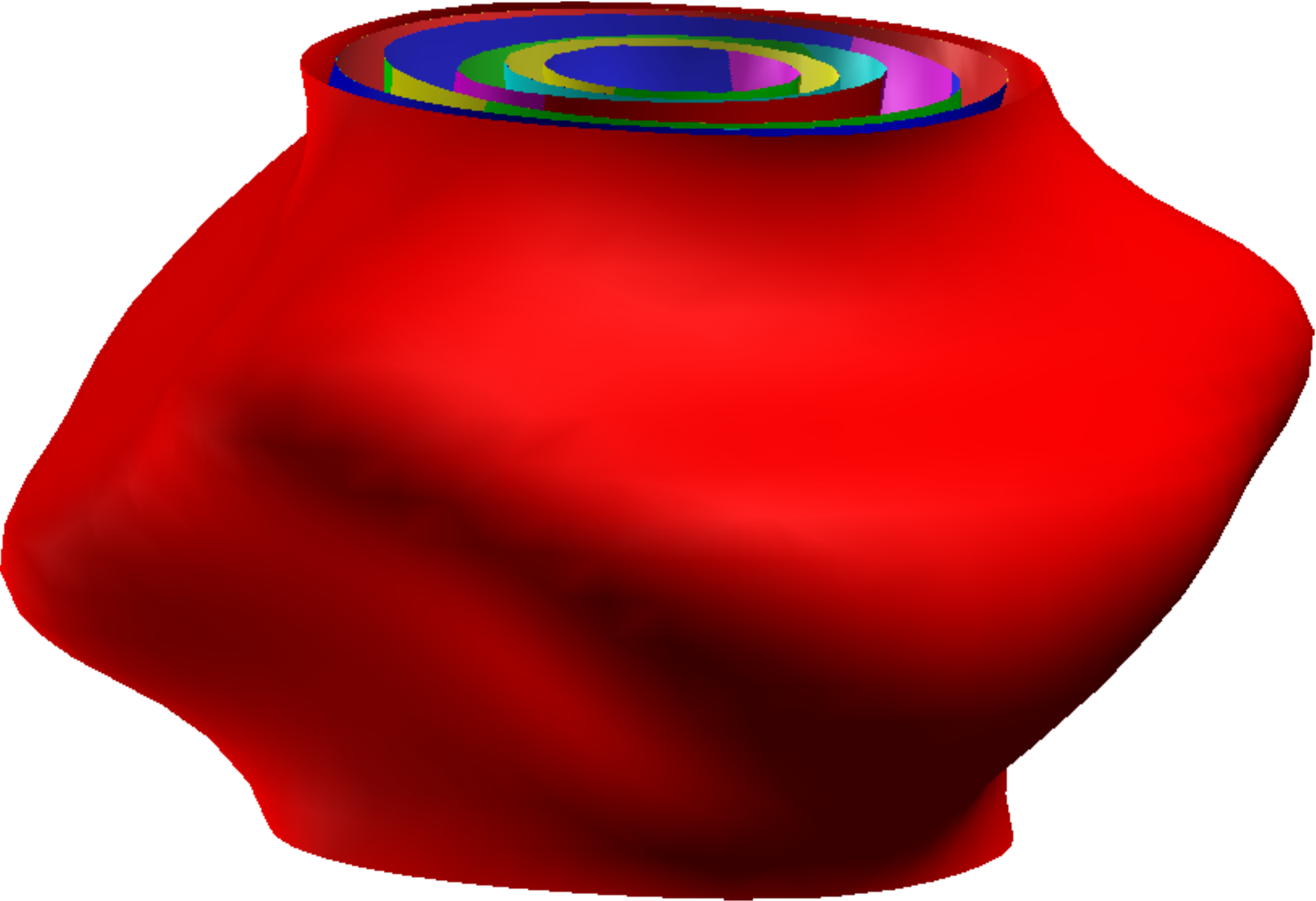}
            \label{fig:Hf_ls}
        }
    \end{center}
        \caption{(color-online) Comparison between the Fermi pockets near the $\Gamma$ point obtained in GGA with (right column) and without (left column) spin-orbit coupling for 
the Zr (top row) and Hf (bottow row) based 221 compound. }

        \label{fig:pockets}
\end{figure*}

\begin{figure*}[htb]
    \begin{center}
        \subfigure[]{
            \includegraphics[width=0.45\textwidth]{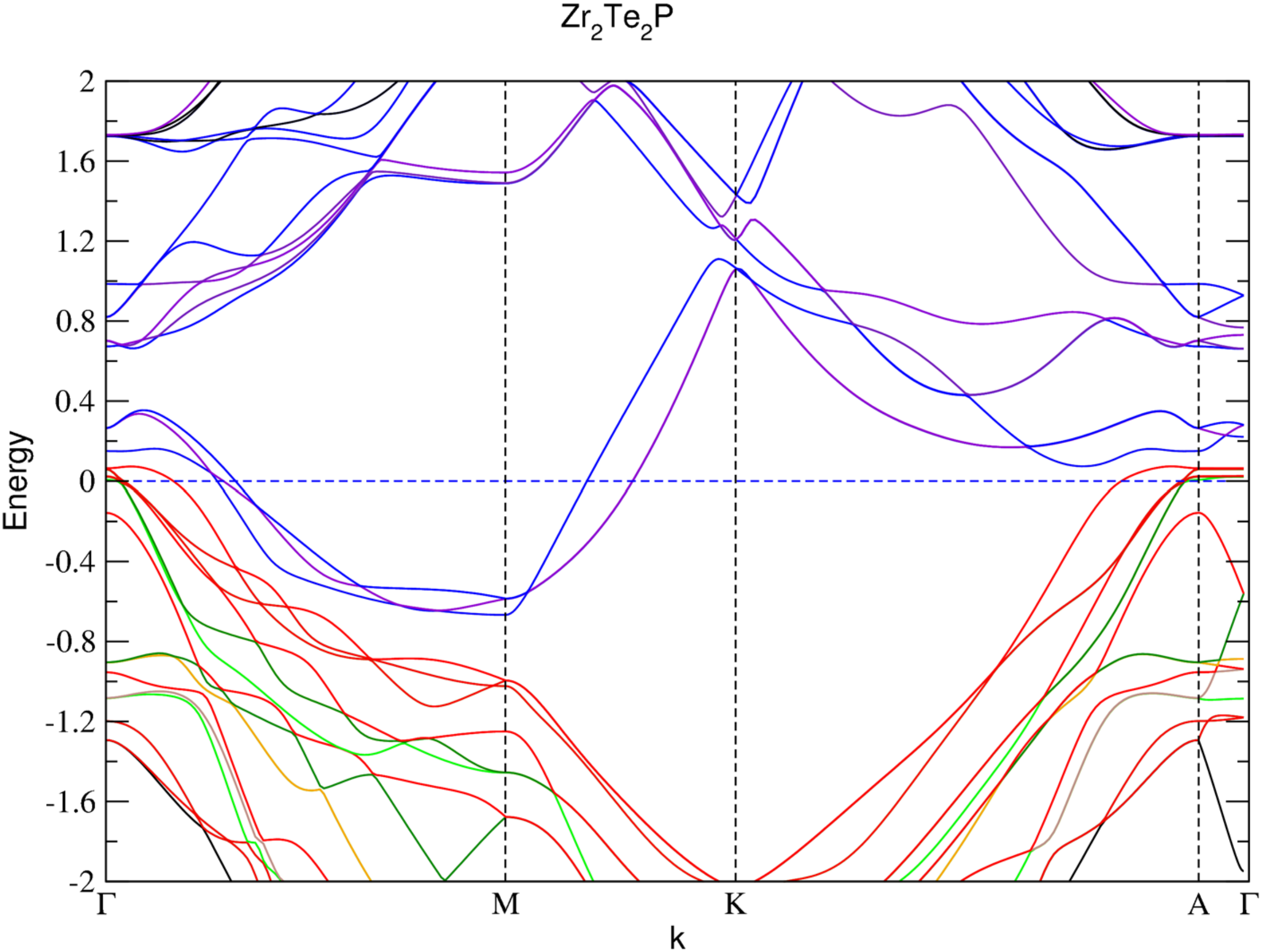}
            \label{fig:Hf_spin}
        } \hskip 0.3 in
        \subfigure[]{
            \includegraphics[width=0.45\textwidth]{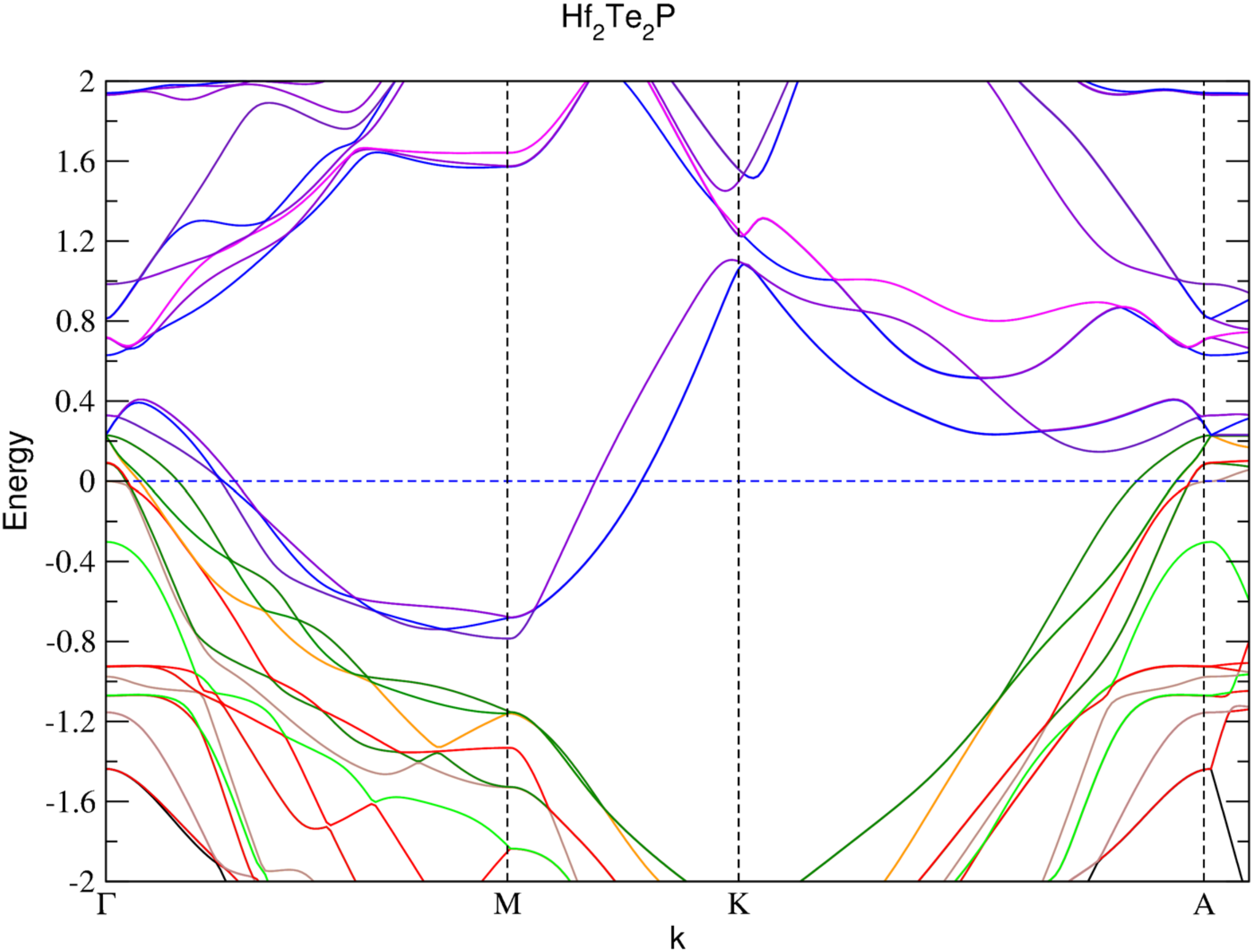}
            \label{fig:Hf_ls}
        }
    \end{center}
        \caption{(color-online)The electronic band structure of (a) the Zr-based   and (b) Hf-based compound obtained with the inclusion 
spin-orbit interaction.}
        \label{fig:Band_structure}
\end{figure*}

The calculated Fermi surfaces and band structures of Zr$_2$Te$_2$P and Hf$_2$Te$_2$P are shown in Fig.~\ref{fig:FS} and Fig.~\ref{fig:Band_structure}, respectively, revealing that they are quite similar. Fig.~\ref{fig:pockets} compares the effect of the spin-orbit interaction on the  pockets near the $\Gamma$ point obtained in GGA with  and without spin-orbit coupling for the Zr (top row) and Hf (bottow row) based 221 compound. Notice that in going from GGA-spin (left column) to GGA-spin+SOC (right column) the size of the orbits are not significantly affected for the Zr based compound. In the case of the Hf based compound, in the absence of SOC the smaller orbit is estimated to be significantly (about a factor of 2) larger than the smallest observed value of the oscillation frequency ($\sim$ 100 T). However, as can be seen from the bottom row of Fig.~\ref{fig:pockets} the effect of SOC is to reduce the lowest orbit by a significant factor which brings the value of the smallest frequency ($\sim$ 80 T) in the observed range. Rough estimates of the size of the three lowest frequencies for the Zr based compound are $\sim$ 30 T, 60 $-$ 70 T, 110 $-$ 120 T. For the case of the Hf based compound the three lowest frequencies are roughly 70 $-$ 80 T, 110 $-$ 120 T, and  $\sim$ 200 T. Fig.~\ref{fig:Band_structure} presents the band structure for the Zr and Hf based compounds obtained with the inclusion of the SOC. They are very similar, where there are small differences in the size of the energy separation between the bands  near Fermi level at the $\Gamma$ point. This is also true at the K point for those bands  closest to the Fermi level. 

\section{Conclusion}
We have used iodine vapor phase transport to grow single crystals of the compounds Zr$_2$Te$_2$P and Hf$_2$Te$_2$P. Measurements of the bulk electrical transport and thermodynamic properties indicate Fermi liquid behavior at low $T$. Both compounds readily exhibit quantum oscillations in magnetization measurements, from which we have determined the frequencies associated with several parts of the Fermi surfaces. Lifshitz-Kosevich fits to the temperature dependent amplitudes of the oscillations reveals small effective masses, with a particularly small value for the $\alpha$ branch of Zr$_2$Te$_2$P. Electronic structure calculations are presented that are in good agreement with these results. Calculations also indicate that measurements in higher magnetic fields will uncover larger Fermi surface sections with higher oscillation frequencies. Fits to the quantum oscillation data also provide preliminarly evidence that some bands have a Berry's phase. This will be presented in detail in a separate manuscript.

These results make both Zr$_2$Te$_2$P and Hf$_2$Te$_2$P intriguing candidates for further study as potential hosts for Dirac physics. Of particular interest is to determine whether the calculated Dirac points that are away from the Fermi energy have an impact on the physical properties, or if the Fermi energy can be tuned to be closer to them. These materials are also particularly well suited to spectroscopic investigations such as ARPES, given the ease with which they are cleaved. This property also opens the possibility of device development. Finally, we point out that this family of materials likely can be expanded to include a variety of chemical analogues: e.g., $M$$_2$Te$_2$As ($M$ $=$ Ti, Zr, and Hf) are likely to form.

\section{Acknowledgements}
This work was performed at the National High Magnetic Field Laboratory (NHMFL), which is supported by National Science Foundation Cooperative Agreement No. DMR-1157490, the State of Florida and the DOE. A portion of this work was supported by the  NHMFL User Collaboration Grant Program (UCGP). LB, DR, and SM are supported by DOE-BES through award DE-SC0002613. TB and TS are supported by DOE-BES through award DE-SC0008832.

\newpage
\bibliography{bib,221cite}

\end{document}